# THE ORIGIN OF INTERMEDIATE-REDSHIFT Lyα ABSORPTION SYSTEMS

KENNETH M. LANZETTA

*Astronomy Program, Department of Earth and Space Sciences*
*State University of New York at Stony Brook, Stony Brook, NY 11794-2100, USA*

DAVID V. BOWEN

*Space Telescope Science Institute, 3700 San Martin Dr., Baltimore, MD 21218, USA*

DAVID TYTLER

*Center for Astrophysics and Space Sciences, University of California, San Diego*
*CASS 0111, 9500 Gilman Dr., La Jolla, CA 92093-0111, USA.*

JOHN K. WEBB and KATRINA M. SEALEY

*School of Physics, University of NSW, P.O. Box 1, Kensington, 2033, NSW, Australia*

## ABSTRACT

We present initial results of an imaging and spectroscopic survey of faint galaxies in fields of *Hubble Space Telescope* spectroscopic target QSOs. The primary objectives of the survey are (*a*) to determine the incidence, extent, and covering factor of extended gaseous envelopes of luminous galaxies and (*b*) to determine the fraction of Lyα absorption systems that arise in luminous galaxies. The goal of the survey is to identify in each field under consideration all objects with apparent $r$-band magnitudes satisfying $r < 21.5$ within angular distances to the QSOs satisfying $\theta < 1.3$ arcmin. We find that at $z \lesssim 1$ most luminous galaxies are surrounded by extended gaseous envelopes of $\approx 160\ h^{-1}$ kpc radius and roughly unit covering factor and that at $z \lesssim 1$ the fraction of Lyα absorption systems that arise in luminous galaxies is at least $0.35 \pm 0.10$ and is likely $0.65 \pm 0.18$.

## 1. Introduction

The Lyα absorption systems seen in the spectra of background QSOs (including the "Lyα-forest" absorption systems that show Lyα absorption but no corresponding metal-line absorption) are easily detected at neutral hydrogen column densities as low as $N \approx 2 \times 10^{13}$ cm$^{-2}$ and provide a uniquely sensitive probe of extremely tenuous material. These absorption systems are accessible to ground-based observation only at redshifts $z \gtrsim 1.6$, which is far beyond the realm of galaxies identified in the deepest galaxy redshift surveys. In contrast, these absorption systems are now accessible to *Hubble Space Telescope* (*HST*) observation at just those redshifts $z \approx 0.1 - 0.7$ that galaxies are routinely identified in deep galaxy redshift surveys[3][4]. This new possibility of observing galaxies and Lyα absorption systems over a common redshift interval provides an unprecedented opportunity to determine the gaseous extent of galaxies and the origin of Lyα absorption systems.

Here we present initial results of an imaging and spectroscopic survey of faint galaxies in fields of *HST* spectroscopic target QSOs. Distances and apparent and absolute magnitudes are expressed in terms of the dimensionless Hubble constant $h \equiv H_0/(100\ \text{km s}^{-1}\ \text{Mpc}^{-1})$ and are calculated (unless otherwise indicated) assuming a deceleration parameter of $q_0 = 0.5$.

Our approach differs from that of a recent survey [7], which concentrated on the large-scale galaxy environments of absorption systems rather than on galaxies responsible for absorption systems. In that study, bright ($B < 19$) galaxies were identified within large angular distances ($\theta < 1°$) to one QSO, thus that survey is most sensitive to galaxies at relatively large impact parameters ($\rho \lesssim 5\ h^{-1}$ Mpc). In contrast our survey is most sensitive to galaxies at relatively small impact parameters ($\rho \lesssim 300\ h^{-1}$ kpc).

## 2. Observational Goal of the Survey

To fulfill the objectives of the survey it is necessary to identify galaxies in fields of QSOs at impact parameters of up to and beyond the gaseous extent of luminous galaxies.

Lacking direct knowledge of the gaseous extent of luminous galaxies, we instead consider the characteristic galaxy absorbing radius required to explain the observed rate of incidence of absorption systems. Adopting the standard assumptions that all absorption systems arise in galaxies and that galaxies are fixed in comoving coordinates and taking the value $n(0.35) \approx 21$ appropriate for Ly$\alpha$ absorption systems at the redshift $z = 0.35$ typical of those detected by *HST* [1], the characteristic absorbing radius need to explain the observed rate of incidence of Ly$\alpha$ absorption systems may be written

$$R_* \approx 320(\epsilon\kappa)^{-1/2}\ h^{-1}\ \text{kpc}. \tag{1}$$

We therefore expect that to fulfill the primary objectives of the survey it is necessary to identify galaxies in fields of QSOs at impact parameters of up to several hundred $h^{-1}$ kpc.

Considering this requirement together with the observational capabilities of current telescopes and detectors, we adopt as the goal of the survey to identify in each field under consideration *all* objects (including stars, compact galaxies, and diffuse galaxies) with apparent $r$-band magnitudes satisfying

$$r < 21.5 \tag{2}$$

within angular distances to the QSOs satisfying

$$\theta < 1.3\ \text{arcmin}. \tag{3}$$

Fields are selected without regard to *a priori* knowledge of the presence or absence of absorption systems. The corresponding luminosity and impact parameter thresholds are shown as functions of redshift in Figures 1 and 2, respectively, from which it is seen that at the redshifts $0.1 \lesssim z \lesssim 0.6$ typical of galaxies with $r < 21.5$ [3,4] the conditions of equations (2) and (3) result in sensitivity to galaxies with luminosities greater than $\approx 0.01 - 1.0\ L_*$ and with impact parameters less than $\approx 80 - 320\ h^{-1}$ kpc. For the faint galaxies considered here $r = 21.5$ corresponds to $B \approx 23.0$ [3], so the condition of equation (2) is comparable to that of other recent deep redshift surveys.

Although the task of identifying *all* objects satisfying equations (2) and (3) is observationally demanding, we consider it essential for several reasons. First, it assures a sample of galaxies which is unbiased with respect to absorption and which contains both galaxies that *do* and galaxies that *do not* give rise to absorption systems, which is crucial to fulfilling the objectives of the survey. Second, it assures that galaxies are not accidentally confused with stars, which is essential for deriving a complete sample of galaxies. Finally, it assures that spectroscopic observations of a number of stars are obtained in each field, which is useful for quantifying the measurement uncertainties of the observations.

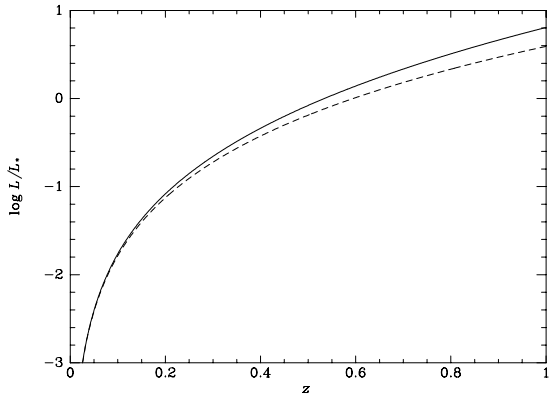 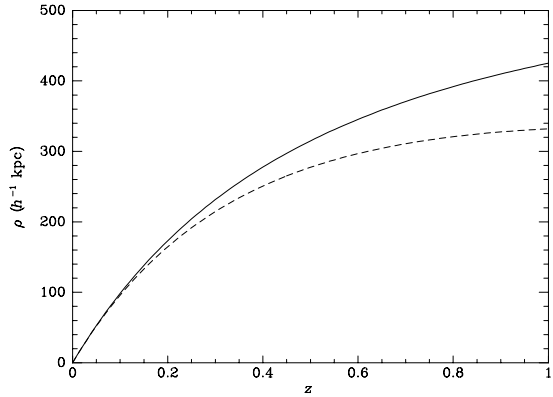

Fig. 1.— Luminosity threshold $\log L/L_*$ vs. redshift $z$ for apparent $r$-band magnitude threshold $r = 21.5$. Calculated assuming $M_{B_*} - 5\log h \approx -19.5$ and $M_B - M_r \approx 1.0$ at $z = 0$ and that the spectral energy distribution varies as a function of frequency $\nu$ as $L_\nu \propto \nu^{-2}$ and hence that the $K$ correction varies as $K = 2.5\log(1+z)$ [8]. Solid line is for $q_0 = 0.5$ and dashed line is for $q_0 = 0$.

Fig. 2.— Impact parameter threshold $\rho$ vs. redshift $z$ for an angular distance threshold $\theta = 1.3$ arcmin. Solid line is for $q_0 = 0$ and dashed line is for $q_0 = 0$.

## 3. Analysis

The imaging observations were obtained with the INT, Lick 3.0 m telescope, and the AAT. The spectroscopic observations were obtained with the KPNO 4 m telescope and the Lick 3.0 m telescope. Galaxies were identified from the spectroscopic observations on the basis of [O II] $\lambda3727$, [O III] $\lambda\lambda4959, 5007$, H$\alpha$, and [S II] $\lambda6724$ emission lines, Fe I, Ca II H and K, G-band, Mg I, Na I, and Balmer absorption lines, and the "4000 Å" continuum break. Redshifts were measured by means of Gaussian fits to [O II] $\lambda3727$, [O III] $\lambda\lambda4959, 5007$, H$\alpha$, and H$\beta$ emission lines and to Ca II H and K absorption lines (and for certain stars to Mg I $\lambda5179$ absorption lines). Measurement uncertainties in the redshifts were estimated to be $\approx 0.0005$, based on repeatability of the individual fits and dispersion of measured stellar velocities. Absorption systems were identified from the $HST$ spectra and absorption line lists[1] and from similar $HST$ spectra obtained from the $HST$ archive.

The current observations cover six fields and are 37% complete to the goal of the survey. These observations identify 46 galaxies at redshifts spanning $z = 0.0700 - 0.5526$ and at impact parameters to the QSOs spanning $\rho = 16.6 - 346.9$ $h^{-1}$ kpc. Of these galaxies, 11 are coincident in redshift with absorption systems and 21 do not give rise to absorption to within sensitive upper limits. Nine galaxies are coincident in redshift with "Ly$\alpha$-forest" absorption systems that show Ly$\alpha$ absorption but no corresponding metal-line absorption, and two galaxies are coincident in redshift with C IV absorption systems that show both Lyα and C IV absorption.

Before the objectives of the survey can be fulfilled it is necessary to establish the relationship between the coincident galaxies and absorption systems. That it, it is necessary to determine whether (*a*) the galaxies are present as the result of chance coincidence, (*b*) the galaxies are merely spatially correlated with the absorption systems, or (*c*) the galaxies are responsible for the absorption systems.

*3.1. Relationship between the Coincident Galaxies and Absorption Systems*

Are the coincident galaxies present as the result of chance coincidence, or are the galaxies physically associated with the corresponding absorption systems?

To distinguish between these possibilities we consider the probability of obtaining the observed redshift agreements between the coincident galaxies and absorption systems by chance. Figure 3 shows the velocity pairs distribution of the galaxies and absorption systems of the survey. It is apparent from Figure 3 that the velocity pairs distribution shows a significant excess at velocity separations $\Delta v < 250$ km s$^{-1}$. The mean background level determined at $\Delta v = 1000 - 10,000$ km s$^{-1}$ is 0.92 pairs per 250 km s$^{-1}$ velocity bin, which yields a Poisson probability of detecting the eight pairs observed at $\Delta v < 250$ km s$^{-1}$ of just $5.1 \times 10^{-6}$ and excludes at the 95% confidence level that more than three of the eight pairs observed at $\Delta v < 250$ km s$^{-1}$ occur by chance. This result demonstrates that the probability of obtaining the observed redshift agreements between the coincident galaxies and absorption systems by chance is negligibly small.

We therefore conclude that most of the coincident galaxies are not present as the result of chance coincidence but rather are physically associated with the corresponding absorption systems.

Are the coincident galaxies merely spatially correlated with the corresponding absorption systems, or are the galaxies responsible for the absorption systems?

To distinguish between these possibilities we first consider the relative velocities between the coincident galaxies and absorption systems. Figure 4 shows in the top panel the relative velocity distribution of the coincident galaxies and absorption systems of the survey and in the bottom panel the velocity distribution of stars measured from the same data set. This latter distribution represents the intrinsic velocity distribution of Galactic halo stars convolved with the measurement uncertainties of the observations and hence provides an appropriate basis for comparison with the relative velocity distribution. It is apparent from Figure 4 that the relative velocity distribution is statistically indistinguishable from the stellar velocity distribution measured from the same data set. Both distributions are characterized by similar means (26 km s$^{-1}$ and $-52$ km s$^{-1}$, respectively) and similar standard deviations (189 km s$^{-1}$ and 140 km s$^{-1}$, respectively), and application of a Kolmogorov-Smirnov test fails to reveal a statistically significant difference between the distributions. This result demonstrates that the relative velocities between the coincident galaxies and absorption systems span the range $\lesssim 250$ km s$^{-1}$ characteristic of motions within individual galaxies rather than the range $\approx 200 - 400$ km s$^{-1}$ characteristic of motions within groups of galaxies or the range $300 - 1000$ km s$^{-1}$ characteristic of motions within clusters of galaxies.

We next consider the presence or absence of a statistical correlation between physical properties of the galaxies and physical properties of the corresponding absorption mea-surements. Figure 5 shows the dependence of Ly$\alpha$ rest-frame equivalent width on impact parameter, and Figure 6 shows the dependence of C IV rest-frame equivalent width on impact parameter. It is apparent from Figure 5 that there exists a statistical anti-correlation between Ly$\alpha$ rest-frame equivalent width and impact parameter. The generalized Kendall rank correlation test indicates that this anti-correlation is significant at the 99.9% level. This result suggests a direct physical connection between the coincident galaxies and the corresponding absorption systems and is expected to obtain if the absorption systems arise in gaseous structures that are centered on the galaxies.

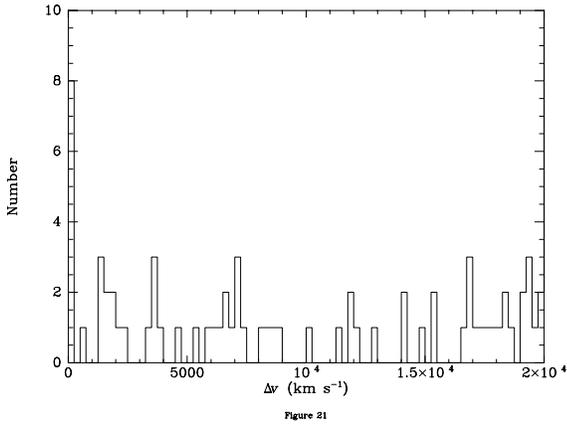 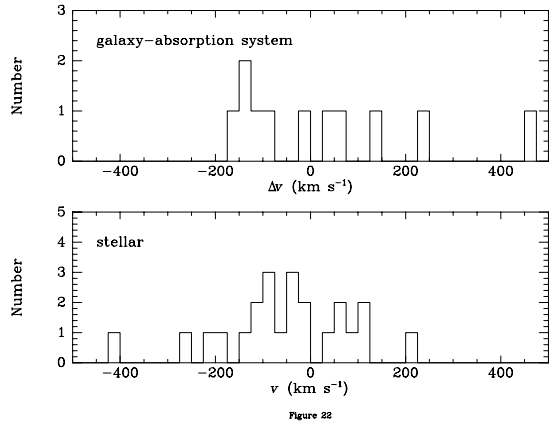

Fig. 3. Velocity pairs distribution of the galaxies and absorption systems of the survey vs. velocity separation $\Delta v$.

Fig. 4. Relative velocity distribution of the coincident galaxies and absorption systems of the survey vs. velocity separation $\Delta v$ (*top*) and stellar velocity distribution vs. velocity $v$ measured from the same data set (*bottom*).

We therefore conclude that the coincident galaxies are not merely spatially correlated with the corresponding absorption systems but rather are *responsible* for the absorption systems.

## 4. Conclusions

Various lines of evidence demonstrate that the coincident galaxies are *responsible for* the corresponding absorption systems and are not present as the result of chance coincidence or merely spatially correlated with the absorption systems. The most important evidence is that there exists a statistical anti-correlation between Ly$\alpha$ rest-frame equivalent width and impact parameter to the QSO. Each of five galaxies with $\rho < 70\ h^{-1}$ kpc gives rise to Ly$\alpha$ absorption, whereas only five of 10 galaxies with $\rho = 70 - 160\ h^{-1}$ kpc give rise to Ly$\alpha$ absorption, and just one of nine galaxies with $\rho > 160\ h^{-1}$ kpc gives rise to Ly$\alpha$ absorption. Furthermore, we find that at least eight of 23 Ly$\alpha$ absorption systems in a homogeneous sample arise in galaxies.

On the basis of these results we reach the following conclusions:

1. At $z \lesssim 1$ most luminous galaxies are surrounded by extended gaseous envelopes of $\approx 160\ h^{-1}$ kpc radius and of roughly unit covering factor. This conclusion confirms previous speculation that normal, luminous galaxies possess extended gaseous halos[2] or extended gaseous disks[5][6].

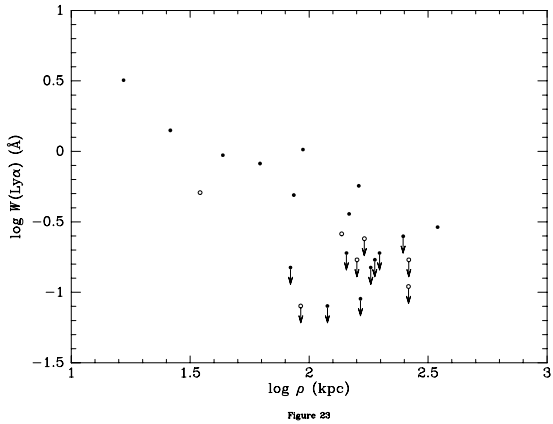

Fig. 5. Log of Lyα rest-frame equivalent width $W(\text{Ly}\alpha)$ vs. impact parameter $\rho$.

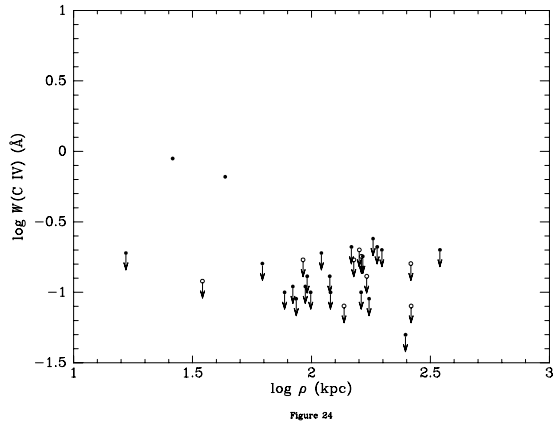

Fig. 6. Log of C IV λ1548 equivalent width $W(\text{C IV})$ vs. impact parameter $\rho$.

2. At $z \lesssim 1$ the extended gaseous envelopes that surround most luminous galaxies do not produce C IV beyond $\approx 50\ h^{-1}$ kpc radius.

3. At $z \lesssim 1$ the fraction of Lyα absorption systems—including Lyα-forest absorption systems—that arise in galaxies is at least 0.35±0.10. Allowing for the known incompleteness of the survey, the actual fraction is likely 0.65 ± 0.18. These conclusions run contrary to the longstanding belief that Lyα-forest absorption systems arise in intergalactic clouds[9].

The work described in this paper is the summary of a more detailed description which was submitted to ApJ in December 1993 (Lanzetta, Bowen, Tytler, Webb).